\documentclass[preprint,prd]{revtex4}
 
\usepackage{amsbsy}
\usepackage{amssymb}
\usepackage[sumlimits]{amsmath}
\usepackage{graphicx}
\usepackage[active]{srcltx}
\usepackage{pdfsync}

\usepackage{color}

\usepackage{cancel}

\def\X{{\mathrm{x}}}
\def\Y{{\mathrm{y}}}
\def\x{{\mathrm{x}}}
\def\y{{\mathrm{y}}}

\def\n{{\mathrm{n}}}
\def\b{{\mathrm{b}}}

\def\s{{\mathrm{s}}}

\def\n{{\rm n}}
\def\p{{\rm p}}
\def\e{{\rm e}}

\def\s{\mathrm{s}}

\newcommand{\A}{{\cal A}}
\newcommand{\B}{{\cal B}}

\def\n{{\rm n}}
\def\p{{\rm p}}
\def\e{{\rm e}}

\def\e{{\rm e}}
\def\s{{\rm s}}

\def\be{\begin{equation}}
\def\ee{\end{equation}}
\def\beq{\begin{equation}}
\def\eeq{\end{equation}}
\def\bea{\begin{eqnarray}}
\def\eea{\end{eqnarray}}
\def\bear{\begin{eqnarray}}
\def\eear{\end{eqnarray}}

\begin{document}

\title{Beyond ideal magnetohydrodynamics: Resistive, reactive and relativistic plasmas}

\author{N. Andersson$^1$, K. Dionysopoulou$^1$, I. Hawke$^1$ and G.L. Comer$^2$}

\affiliation{
$^1$ Mathematical Sciences and STAG Research Centre, University of Southampton,
Southampton SO17 1BJ, United Kingdom\\
$^2$ Department of Physics, Saint Louis University, St. Louis, MO, 63156-0907, USA}

\begin{abstract}
We develop a new framework for the modelling of charged fluid dynamics in general relativity. The model, which builds on a recently developed variational multi-fluid model for dissipative fluids, accounts for relevant effects like the inertia of both charge currents and heat and, for mature systems, the decoupling of superfluid components. We  discuss how the model compares to standard relativistic magnetohydronamics and consider the connection between the fluid dynamics, the microphysics and the underlying equation of state. As illustrations of the formalism, we consider three  distinct two-fluid models describing i) an Ohm's law for resistive charged flows, ii) a relativistic heat equation, and iii) an equation representing the momentum of a decoupled superfluid component.  As a more complex example, we  also formulate a three-fluid model which demonstrates the thermo-electric effect. This framework allows us to model neutron stars (and related systems) at a hierarchy of increasingly complex levels, and should enable us to make progress on a range of  exciting problems in astrophysics and cosmology.
\end{abstract}

\maketitle

\section{Introduction}

Despite nearly five decades of observations, neutron stars continue to challenge our understanding. We do not (completely) understand why pulsars pulse and we do not (exactly) know why magnetars flare. The origin and evolution of the magnetic field of these systems remain vexing issues. Neutron stars may  host the strongest magnetic fields in the universe, but we have few quantitative models that explain the observed phenomenology. 

The problem is immensely challenging from the theoretical point-of-view. Neutron star modelling involves many extremes of physics, from the composition and state of matter at supranuclear densities and the dynamics of large scale superfluid/superconducting condensates to issues involving the star's nuclear crust and magnetospheric processes. Because of the vastly different length- and timescales involved it is not realistic to expect a theoretical model to cover all aspects. This means that modelling involves judicious choices of what is important in a given situation and what is not. Again, the electromagnetic field provides a good example. So far, following mainstream astrophysics, most studies of the neutron star magnetic field have been based on ideal magnetohydrodynamics. This makes sense because such models are tractable and one can argue that the high conductivity in the star's core supports the involved assumptions. Of course, we also know that there are situations where this model is not appropriate. The obvious example involves a mature star in which the core is cold enough that the protons form a superconductor. In this case, the magnetic field will be carried by quantised fluxtubes and the dynamics of the system differs significantly from ideal magnetohydrodynamics \cite{supercon,gusa}. Similarly, non-ideal effects are key for any problem that involves magnetic field evolution (see \cite{pons1,pons2, gour} for recent discussions). In order to understand the field evolution we need to understand how the resistivity enters (see \cite{namhd} for discussion and references to the literature), whether various ``battery'' terms are important etcetera.

In order to build a new generation of models for neutron star magnetism we have to proceed beyond ideal magnetohydrodynamics. We need to model the resistivity, while at the same time accounting for nuclear reactions and changes to the state of matter in the star's core, e.g. the onset of superfluidity. We need to be able to quantify how the evolution of the magnetic field strains the star's crust and establish whether this allows the build-up of the energy required to power magnetar flares \cite{pons,lander}. Most importantly, we need to be able to model the various scenarios within general relativity. Otherwise, we will not be able to make use of realistic matter models. We will also not be able to test our models against precision observations. 

In this paper we lay the foundation for a new state-of-the-art. Building on the formal results from a variational analysis \cite{variate}, we develop a fully relativistic framework that accounts for four (fluid) components, allowing us to consider the charge current, heat flow and superfluid dynamics, relative to a given bulk flow. The model is designed to make contact with ideal magnetohydrodynamics in the appropriate limit and we discuss a hierarchy of models in order to explore new effects that enter as the problem becomes more complex. Key to developing the framework is the introduction of a suitable family of observers associated with the fluid flow. The introduction of these observers essentially provide a fibration of spacetime. In essence, this leads to a formulation similar to that often used in cosmology (see \cite{tsagas} for a relevant discussion). From the fluid dynamics point-of-view we have a local description and a key part of our argument is that this local frame is required to make contact with the microphysics encoded in the equation of state. The model we develop will not, however, be suitable for numerical simulations of (say) merging neutron stars. To make progress in that direction, we need to connect the present discussion with a 3+1 foliation of spacetime (see, for example, \cite{baum} for a relevant review). This basically involves introducing a different set of observers and clarifying  some additional issues. We develop the required foliation model in a companion paper \cite{foliate}. 

\section{The variational multi-fluid model}

Following \cite{livrev,monster} we take the variational approach to relativistic fluid dynamics as our starting point. This is natural because the formalism is readily generalised to account for individual fluid components with distinct flows. Moreover, it is already established how one includes (at least at the formal level)  both resistivity (eg. friction) and reactions (leading to individual fluxes not being conserved) in this model \cite{variate}.

We take as our starting point the individual particle fluxes $n_\X^a$ in the system, where different fluid components are labelled 
by a constituent index $\X,\Y,\ldots$. This allows us to define the (co-moving) number densities;
\be
n_\x^2 = - g_{ab} n_\x^a n_\x^b \ , 
\ee
where $g_{ab}$ is the (dynamical) spacetime metric, and hence the individual four velocitites, such that;
\be
n_\x^a = n_\x u_\x^a \ .
\ee
Note that the usual summation 
convention applies to the spacetime indices $a,b,c,\ldots$. It does not apply to the constituent indices $\x,\y,\ldots$. 

We consider a four-component system composed of neutrons (n), protons (p), electrons (e) and entropy (s). This is the minimum level of complexity required if we want to consider realistic neutron star dynamics. The relative flow of the protons and electrons  leads to the charge current that couples the material motion to electromagnetism. The entropy flow is key if we want to account for the redistribution of heat, which we need to track if we want to consider (say) the cooling of a young neutron star. Finally, the neutrons need to be accounted for because they make up the bulk of the star. Moreover, as the star matures  the neutrons  become superfluid and (at least partially) decouple from the other components. In order to explore the evolution and dynamics of maturing neutron stars, we need to allow for the relative flows of these different components. 

In order to extend the model beyond the two-component case that was considered in \cite{namhd}, we  first of all do not assume that the individual fluxes are conserved. In general, we then have
\begin{equation} 
    \nabla_{a} n^{a}_\X = \Gamma_\X \ , \label{consv2} 
\end{equation} 
where $\Gamma_\X$ is the relevant creation/destruction rate. The presence of reactions impacts on the momentum equations for the fluids (via what would be ``rocket'' terms in Newtonian physics). If we allow (at least some of) the components to be charged and account for resistivity, the variational approach leads to \cite{variate}
\be
2 n_\x^b  \nabla_{[b} \tilde \mu^\x_{a]} + \Gamma_\x \tilde \mu^\X_a =  R^\X_a \ ,
\label{momeq}
\ee
where the square brackets indicate anti-symmetrisation and we have defined
\be
\tilde \mu^\x_a = \mu^\x_a + e_\x A_a \ ,
\ee
with $e_\x$  the electric charge of the $\x$-component and $A_a$ is the electromagnetic vector potential. 

The resistivity must satisfy the constraint
\be
\sum_\x R^\x_a = 0 \ .
\label{Rcon}
\ee
We see that, by contracting \eqref{momeq} with  $n_\X^a$, and introducing the chemical potential for each component as $\tilde \mu_\x = -u^a_\x \tilde \mu^\x_a$, we must have
\be
\Gamma_\x = - {1\over \tilde \mu_\x} \left( u_\x^a R^\x_a \right) \ . 
\label{gamdef}
\ee
That is, the reaction rate determines the time component (in a co-moving frame) of the resistivity.

As usual \cite{livrev}, the fluid part of the conjugate momentum, $\mu^\x_a$, for each component follows from an energy functional $\Lambda$  (which encodes the microphysics from the equation of state, but which does not account for the coupling to electromagnetism);
\begin{equation} 
    \mu^\X_{a} = {\partial \Lambda \over \partial n_\x^a} =  g_{ab} \left(\B^{\X } n^b_\X + \sum_{\y\neq\x}\A^{\X \Y} 
                   n^b_\Y\right) \ , \label{mudef2} 
\end{equation} 
where
\be
\B^\x = -  2 \frac{\partial \Lambda}{\partial   n^2_{\X}} \ , 
\ee
and 
\be 
    \A^{\X \Y} = \A^{\Y \X} = - \frac{\partial \Lambda}{\partial 
                 n^2_{\X \Y}} \quad , \quad \X \neq \Y \ . \label{coef12} 
\ee 
Basically, we need to consider both the (co-moving) number density $n_\x$ from above and
\be
n_{\x\y}^2 = - n_\x^a n^\y_a \ , \qquad \y \neq \x \ .
\ee
The  $\A^{\X \Y}$ coefficients represent the fact that each fluid momentum  $\mu^\X_{a}$ 
may, in general, be given by a linear combination of the different  $n^{a}_\X$
currents.  This is  usually referred to as the entrainment effect. In the problems we are (mainly) interested in the effect is important for two reasons. First, the strong interaction leads to a moving neutron being endowed with a virtual cloud of protons (and vice versa). This means that it may be more difficult  (or indeed, easier) to move the nucleons through the mixture than one might have expected (see for example \cite{alpar}). 
As discussed in \cite{namhd,heat1,heat2} it is also important to consider entrainment between material particles and the entropy. This is required to ensure causality of the heat flow. We clearly need to account for these two mechanisms. We are, however, not aware of any argument that suggests entrainment between leptons and baryons so will not consider this possibility here. 

Finally, in order to account for the coupling between the matter flow and the dynamics of spacetime, we need the matter stress-energy tensor for the multifluid system;
\be
     T^{ab}_\mathrm{M} = \Psi g^{ab}+ \sum_{\X} 
                       n^a_\X \mu_\X^b \ , \label{seten2} 
\ee
where we have introduced the generalized pressure $\Psi$ as
\begin{equation}
     \Psi = \Lambda - \sum_{\X } n^{a}_\X \mu^\X_{a} \ . 
\end{equation} 

The (minimal) coupling to electromagnetism is effected in the same way as in \cite{namhd}. 
The  electromagnetic Lagrangian is built from the anti-symmetric Faraday tensor;
\be
F_{ab} = 2 \nabla_{[a} A_{b]}  \ ,
\ee 
and the  electromagnetic field couples to the matter flow through the charge current $j^a$. In order for this construction to be gauge invariant, we must insist that the current is 
conserved. That is, we have the constraint
\be
\nabla_a j^a = 0  \ .
\ee
In the variational model, the charge current is given by the sum
\be
j^a = \sum j_\x^a = \sum_\x e_\x n_\x^a \ .
\ee
In the  case of conserved fluxes \cite{namhd}, the gauge constraint is automatically satisfied. When we account for reactions, this is no longer the case and we must impose charge conservation
\be
\sum_\x e_\x \Gamma_\x = 0 \ .
\label{gaugecon}
\ee

As usual, variation of the electromagnetic contribution to the Lagrangian with respect to the vector potential (keeping $j^a$ fixed!), leads to the Maxwell equations
\be
\nabla_b F^{ab} = \mu_0 j^a \ ,
\label{max1}
\ee
where $\mu_0$ is the relevant coupling constant, which are completed by 
\be
\nabla_{[a}F_{bc]}=0 \ .
\label{max2}
\ee
The latter equation is automatically satisfied given the anti-symmetry of  $F_{ab}$.

A variation with respect to the spacetime metric leads to the electromagnetic contribution to the stress-energy tensor being given by
\be
T_{ab}^\mathrm{EM} =  {1\over \mu_0} \left[g^{cd}F_{ac}F_{bd}-{1\over 4}g_{ab}\left( F_{cd}F^{cd}\right) \right] \ ,
\label{TEM}\ee
where
\be
\nabla_a T^{ab}_\mathrm{EM} =  j_a F^{ab} \equiv - f_\mathrm{L}^b \ ,
\label{loren}\ee
 defines the Lorentz force $f_\mathrm{L}^a$. Identifying the individual contributions to the Lorentz force, we can rewrite \eqref{momeq} as
\be
2 n_\x^b  \nabla_{[b} \mu^\x_{a]}  + \Gamma_\x  \mu^\x_a =  j_\x^b F_{ab}+ R^\X_a - \Gamma_\x e_\x A_a \ .
\label{momeqs}
\ee

Adding these equations, and recalling the constraint \eqref{Rcon}, 
\be
\nabla_a T^{ab}_\mathrm{M} = -  \nabla_a T^{ab}_\mathrm{EM} = - j_a F^{ab} = f_\mathrm{L}^b \ ,
\ee
where we have the total stress-energy tensor 
\beq
  T^{ab}=T^{ab}_\mathrm{M}+T^{ab}_\mathrm{EM} \ .
\eeq
As in the case of non-dissipative flows, it is easy to show that, for a solution to the combined 
 fluid equations,  \eqref{consv2} and \eqref{momeqs},  and Maxwell's equations it is 
automatically true that $\nabla_{a} T^{ab} = 0$. 

\section{Making contact with ``standard'' MHD}

As we develop a more realistic model for relativistic magnetohydrodynamics, it is important to keep in mind the intended applications and potential numerical simulations. 
In principle, the model outlined in the previous section provides a complete (once we provide an equation of state and the relevant microphysics information) description of the charged multifluid dynamics. However,  this framework includes a number of aspects which are not included in current state-of-the-art simulations. From a practical point-of-view, it would not make sense to try to account for all these aspects in one go. As we consider a higher level of realism, we also have to be realistic. Multifluid simulations, with the number of components we envisage, are likely to be costly. Another reason to be cautious is the fact that some aspects may not be within reach of nonlinear simulations at all, simply due to a mismatch of timescales. It is important to keep these caveats in mind as we proceed.

The ultimate aim of this work is to build a framework that accounts for the constituents that make up the outer core of a realistic neutron star; neutrons, protons, electrons and entropy (heat). Given the equations from the previous section, the next natural step would be to make contact with the way the problem is usually described. As a first step in this direction, let us try to connect our general framework to the standard formulation for relativistic magnetohydrodynamics. 

Almost exclusively, simulations involving relativistic magnetohydrodynamics take as their starting point baryon number conservation:
\be
\nabla_a \left( n u^a \right) = 0 \ ,
\label{cons}\ee
combined with the  standard perfect fluid stress-energy tensor:
\be
T_\mathrm{M}^{ab} = \left( p+\varepsilon\right) u^a u^b + p g^{ab}  \ ,
\label{Tab}\ee
where $p$ is the pressure and $\varepsilon$ is the energy density.
These equations tend to be ``assumed'' rather than derived. In Newtonian theory, the corresponding results can be obtained from multi-component plasma equations analogous to \eqref{consv2} and \eqref{momeqs} \cite{kulsrud}. The relations \eqref{cons} and \eqref{Tab} would be the direct generalisation of the non-relativistic results. However, one should perhaps exercise some caution because the elevation to relativity may not be this obvious. A particular issue that comes into play is the choice of observer. To be specific: Does there exist an observer, with four velocity $u^a$, such that both \eqref{cons} and \eqref{Tab} are true based on the multifluid model? As we will now demonstrate, the answer is (strictly speaking) no.

Another reason to explore the validity of \eqref{cons} and \eqref{Tab} for the multicomponent problem relates to the desire
to describe more realistic astrophysical systems. A particular illustration of this involves efforts to develop ``non-ideal'' magnetohydrodynamics, e.g., by accounting for resistivity. Recent efforts in this direction have adopted what one might perhaps call a ``bottom up'' approach, following \cite{beke}, where one takes  ideal magnetohydrodynamics as a starting point and adds in a phenomenological resistivity in what seems a ``natural'' way.
This involves adding terms that are expected to be small, under most circumstances, to the equations and tracking the effects  these changes have on the dynamics. This is a reasonable procedure, but it raises another important question: How do we know that the small terms that we add back in are more important than  small terms we threw away when we reduced the problem to ideal magnetohydrodynamics in the first place? Without quantifying how the ideal equations deviate from a higher-level model, we can not answer this question. 

It is obvious from the outset that the general multifluid description (or indeed the corresponding kinetic description), that keeps track of the individual fluxes, can not lead to ``single-fluid'' equations like \eqref{cons} and \eqref{Tab} unless we make simplifying assumptions. We need to understand what these assumptions are and under what circumstances they make sense.
To make the problem precise, let us focus on a system with two charged components, each carrying a single  unit of charge. Since we have neutron stars in mind, these components would be the protons (p) with $e_\p=e$ and the electrons (e) with $e_\e=-e$. The individual  charge currents are then
$j_\p^a=en_\p^a$ and $j_\e^a=-en_\e^a$ and we see that the gauge constraint \eqref{gaugecon} means that we must have
\be
\Gamma_\p = \Gamma_\e \ .
\label{gpe}
\ee
In addition, we have the neutrons (n) which are, of course, charge neutral. As far as electromagnetism is concerned the neutrons are passive bystanders, but they play a key role in the fluid dynamics. The heat is accounted for in terms of the entropy (s), and we will assume that the mean free path of the associated phonons etcetera is short enough that the entropy can also be considered as a fluid \cite{heat1,heat2}. This assumption restricts the validity of the model somewhat, but it seems like a reasonable starting point. 

In order to make contact between the general relativistic model and the microphysics that determines the matter composition, the reaction rates and so on, we need to choose a suitable observer frame. In principle, the different flows may move at high velocity with respect to this frame, which would necessitate individual component Lorentz factors. However, in many physical settings there will exist a family of observers such that each relative flow is represented by a slow relative drift. (If this is not the case, the problem can definitely not be reduced to an effective single-fluid model.) This means that it makes sense to linearise the flows relative to the observer, which moves with  four velocity $u^a$ (normalised such that $u^au_a=-1$), in such a way that we have: 
\be
u_\x^a = \gamma_\x \left(u^a + v_\x^a\right) \ , \qquad \mbox{where} \qquad u^a v^\x_a = 0 \ , \qquad \mbox{and} \qquad \gamma_\x = \left( 1 - v_\x^2 \right)^{-1/2} \ ,
\ee
with the ``drift'' velocities $v_\x^a$  small enough that 
$\gamma_\x\approx 1$. Note that the assumption of small drift velocities does not impose any restrictions on the bulk flow associated with $u^a$. 
Within this linear model, let us consider the equation for baryon number conservation \eqref{cons}. Provided we neglect the individual $\gamma_\x$, it is easy to see that all observers would measure the same number densities $n_\x$. This means that they would agree on the baryon number $n=n_\n+n_\p$. Baryon number  conservation then simply corresponds to imposing 
\be
\Gamma_\n+ \Gamma_\p = 0 \ .
\ee
Of course, in the neutron star case the  reaction rates  have to balance as they are due to the Urca reactions; 
\begin{eqnarray}
\p+\e &\rightarrow& \n+\nu_\e \ , \\
\n &\rightarrow& \p+\e+\nu_e \ .
\end{eqnarray}

This argument tells us how the rates which enter into the multifluid formalism follow from the microphysics. The rates depend on the chemical imbalance 
\be
\beta = - u^a \left( \mu^\n_a - \mu^\p_a - \mu^\e_a\right) =  \mu_\n - \mu_\p - \mu_\e \ .
\ee
Again, this should not lead to confusion since $\beta$ is the same according to all observers (in the linear drift model we are considering here).

The connection with the usual ``single-fluid'' conservation law \eqref{cons} is less straightforward.
We would have
\be
\Gamma_\n + \Gamma_\p = \nabla_a ( n_\n u_\n^a + n_\p u_\p^a ) = \nabla_a ( n u^a) + \nabla_a (n_\n v_\n^a + n_\p v_\p^a) = 0  \ .
\ee
In general, the only way to retain \eqref{cons} is to work in a specific observer frame such that~\footnote{Note that the choice of frame is not a gauge constraint. Rather, it involves deciding which observer measures scalar quantities associated with the microphysics, like baryon number density and temperature.}
\be
n_\n v_\n^a + n_\p v_\p^a = 0 \ .
\label{Eckart}
\ee
This is the analogue of the Eckart frame from the problem of relativistic heat flow (see \cite{heat1} for discussion).
However, as we are comparing to  ideal magnetohydrodynamics, it makes sense to assume that the baryons move together. Then we have a single drift velocity $v^a =v_\n^a=v_\p^a$ which would vanish if we choose the co-moving Eckart frame. However, it is easy to see that if we make this choice of frame then \eqref{seten2} is not compatible with the perfect-fluid stress-energy tensor. Even if we lock the heat to the baryons, as well, we  have additional  terms linear in the charge current in the stress-energy tensor. Whether these can be ``ignored'' or not depends on the physical situation.

Having considered the issue of baryon number conservation, let us turn to the corresponding problem for the stress-energy tensor. Is it possible to reduce the multifluid expression \eqref{seten2} to the single-fluid form \eqref{Tab}? The answer is yes, we can (again) do this by choosing an appropriate frame for the observer. We need $u^a$ to be such that that the observer measures no relative energy/momentum flow. This choice would be analogous to the Landau-Lifschitz frame from discussions of relativistic heat flow (again, see \cite{heat1}).

When we ignore the Lorentz factors associated with the individual drift velocities, the matter stress-energy tensor takes the form (note that, in this linear drift model the entrainment terms cancel when we add the components together)
\begin{equation}
T^\mathrm{M}_{ab} = \Psi g_{ab} + \sum_{\x \in \{\n,\p,\e\,\s\}} n_\x \mu_\x  \left( u_a u_b  + v^\x_b u_a+  v^\x_a u_b  \right)  \ .
\end{equation}
Contracting with $u^a$ we obtain an expression for the momentum flux;
\be
u^a T^\mathrm{M}_{ab} = \left( \Psi - \sum_{\x \in \{\n,\p,\e\,\s\}} n_\x \mu_\x   \right) u_b - \sum_{\x \in \{\n,\p,\e\,\s\}}  n_\x \mu_\x   v^\x_b \ .
\label{momflux}\ee
Another contraction, now with $u^b$, leads to the (fluid) energy measured by the observer;
\be
\varepsilon = u^b u^a  T^\mathrm{M}_{ab} = -  \Psi + \sum_{\x \in \{\n,\p,\e\,\s\}} n_\x \mu_\x \ ,
\ee
and (since, to linear order in the drift velocities $\varepsilon= -\Lambda$) we see that it is natural to identify $\Psi$ as the pressure $p$, leading to the standard thermodynamic relation~\footnote{It is worth noting that, in the case of  linear drift velocities, the isotropic pressure also follows from the trace of the stresses. That is, writing  $T^{ab}=\varepsilon u^a u^b + \pi^{ab}$ we have  $p= \pi^a_{\ a}/3$. In the case of nonlinear drift the trace does not give the generalised pressure $\Psi$. }
\be
p + \varepsilon =  \sum_{\x \in \{\n,\p,\e\,\s\}} n_\x \mu_\x \ .
\ee
Alternatively, if we single out the entropy by letting  $n_\s=s$ and note that the chemical potential associated with the entropy is the temperature $T$, we have
\be
p + \varepsilon =  \sum_{\x \in \{\n,\p,\e\}} n_\x \mu_\x + sT \ .
\ee
Returning to the momentum flux \eqref{momflux} we see that, if we choose the observer such that there is no relative momentum flux
\be
\sum_{\x \in \{\n,\p,\e\,\s\}}  n_\x \mu_\x   v^\x_b  =0  \ ,
\label{Landau}
\ee
then we arrive at \eqref{Tab}. Of course, if we work in this frame then baryon number conservation is not given by \eqref{cons}.

The unavoidable conclusion is that there are issues of concern already at the level of ideal magnetohydrodynamics. One would have to, at the very least, check that the deviation from \eqref{cons} and/or \eqref{Tab} do not have an important effect on any given problem.

Interestingly, the problem we have uncovered is not present in one particular (and rather important) case. Consider a two-component pair plasma, with electrons (e) and positrons (p). Then the two chemical potentials are equal, $\mu_\e=\mu_\p$, and it is clearly the case that \eqref{Eckart} and \eqref{Landau} are compatible. In this problem, the single-fluid reduction is safe.

In contrast, suppose we consider a neutron star core and impose \eqref{Landau} together with the assumption that the baryons have a common drift velocity $v^a$. Let us also introduce the charge current. As we have already seen, the  gauge constraint requires
$\Gamma_\p = \Gamma_\e$. 
The upshot of this is that if a fluid element starts out charge neutral then it remains so throughout an evolution. We then have local charge neutrality, $n_\p=n_\e$, which means that 
\beq
j^a = e\left( n_\p - n_\e\right) u^a + e\left( n_\p v_\p^a - n_\e v_\e^a\right) = e n_\e \left( v_\p^a - v_\e^a\right) \equiv J^a \ ,
\eeq
which defines the spatial charge current $J^a$. That is, if we impose charge neutrality then 
\beq
\nabla_a J^a = 0 \ .
\eeq
It is also convenient to introduce  the heat flux
\be
q^a = sT v_\s^a \ .
\ee
In terms of these variables, we find that  the condition from \eqref{Eckart} would be satisfied if
\be
{1\over \mu_\n} \left[ n_\e \beta v^a + {\mu_\e \over e} J^a - q^a\right]=0 \ .
\label{test1}
\ee
This is clearly not true in general, but if the system is cold and in beta equilibrium, then only the term involving the charge current remains. This term is suppressed by the factor $\mu_\e/\mu_\n$, which is small in the Newtonian limit but may be of the order of $0.1$ in a neutron star core. It is easy to envisage situations where this term can be ignored, but it is clear that the model is now becoming contrived. In a general nonlinear situation there is no reason to expect the left-hand side of \eqref{test1} to vanish identically. 

\section{The multifluid model}

The analysis from the previous section provides clear motivation for the multifluid model. Yet it remains the case that the general description may be ``a step too far'' for many relevant applications. Hence, it is natural to discuss simplifications. First of all, let us retain the assumptions of linear drift velocities and local charge neutrality, both of which seem reasonable. Next we make a decision regarding the frame. In the following, we will describe the problem in the Landau-Lifschitz frame \eqref{Landau}. There are two reasons for this decision. First of all, it may be more ``intuitive'' to describe the scattering processes that lead to the resistivity in this frame, as it represents the centre of momentum. Secondly, from a practical point-of-view this choice makes sense. We will outline a set of models in the following and by opting for the Landau-Liftschitz frame we ensure that the form of the equations for total energy and momentum conservation remain the same in all cases. The use of the alternative frame choice will be discussed in the companion paper where the 3+1 foliation view of the multifluid model is developed.

Hence, we introduce an observer such that
\begin{equation}
n_\n \mu_\n v_\n^a + n_\p \mu_\p v_\p^a + n_\e \mu_\e v_\e^a + q^a = 0 \ . \label{eq:observer}
\end{equation}
This means that the fluid stress-energy tensor takes the perfect fluid form \eqref{Tab} and we retain the usual equations for the bulk fluid flow, even when the neutrons are allowed to flow relative to the protons. 

If we want to derive the appropriate form for Ohm's law, we need to work out the equation that governs the evolution of the charge current. Provided the system is charge neutral, we need an equation for $J^a$. Similarly, in order to describe the heat flow, we need an evolution equation for the heat flux $q^a$, and finally, if the neutrons in the system are superfluid then we also need to keep track of their relative flow, $v_\n^a$. The question is if it is possible to keep track of these four degrees of freedom without making the mathematics overwhelming.  

\subsection{Entrainment and effective masses}

An important feature of the multifluid model is the entrainment, essentially a measure of how easy it is for one fluid to flow relative to another. The entrainment enters through the canonical momenta and from the definition \eqref{mudef2} we see that  it leads to the momentum of a given fluid, $\mu_\x^a$, not being aligned with the particle flux, $n_\x^a$. Making use of the definition for the chemical potential, we have (in the linear drift model)
\begin{equation} 
    \mu^\X_{a} =  \mu_\x u_a + \pi^\x_a  \ ,
\end{equation} 
where 
\be
\pi^\x_a = \mu_\x v_a^\x +  \sum_{\y\neq\x}\A^{\X \Y} n_\y w^{\y\x}_a \ , \qquad \mbox{with} \qquad w_{\y\x}^a = v_\y^a - v_\x^a \ .
\ee

The entrainment effect is quite intuitive. It can be expressed in terms of an effective mass for each species. In our discussion, we will account for two different entrainment mechanisms. The first is due to the strong interaction and encodes how each neutron is associated with a virtual cloud of protons, meaning that its inertia  differs from that of a bare neutron (and vice versa). The second entrainment mechanism is associated with the effective inertia of heat, and couples the entropy component to the material components in the system. Accounting for this effect is important, as the associated thermal inertia renders the relativistic model for heat flow causal \cite{heat1,heat2}. This is obviously crucial from a conceptual point of view and it may  be important in practical applications, as well. 

In order to illustrate the link between entrainment and the effective mass, let us consider only the strong interaction induced entrainment between neutrons and protons.  We then have
\be
\pi^\n_a = \mu_\n v_a^\n +  n_\p \A^{\n\p} w^{\p\n}_a \ ,
\ee
and
\be
\pi^\p_a = \mu_\p v_a^\p +  n_\n \A^{\n\p} w^{\n\p}_a \ .
\ee
Considering the first of these relations, we see that the neutron momentum according to someone riding along with the protons (take $v_\p^a=0$) is (this argument is analogous to the Newtonian discussion in \cite{prix})
\be
m_\n^\ast v_\n^a = \left( \mu_\n -n_\p  \A^{\n\p} \right) v_\n^a \ .
\label{neuteff}
\ee
This defines the effective neutron mass $m_\n^\ast$. Conversely, we have
\be
\A^{\n\p} = {1\over n_\p} \left( \mu_\n - m_\n^\ast\right) =  {1\over n_\n} \left( \mu_\p - m_\p^\ast\right)  \ ,
\ee
where we have applied the same argument to arrive at the effective proton mass $m_\p^\ast$. In the general case, with a number of distinct flows (or several entrainment mechanisms), the expression for the effective mass is not as simple as \eqref{neuteff} but the concept still makes sense. 

\subsection{The friction}

We want to build a model that accounts for linear friction, which works to prevent different fluid components from flowing through one another. This is important for conceptual reasons, because it provides a mechanism that allows us to consider the limit where two fluids are locked by strong friction. The inclusion of friction is also central to any non-ideal magnetohydrodynamics model. It is the friction that  leads to both resistivity and thermal conductivity. To make progress we make use of the phenomenological model discussed in \cite{variate}. This model accounts for reactions and resistive scattering and satisfies constraints deduced from the variational analysis. It does not incorporate the many other dissipation channels that may be relevant for a general multi-fluid system (see for example \cite{monster, bryn}). However,  the strategy for  including these mechanisms is relatively clear given the results in \cite{variate} and \cite{monster}. 

Assuming that the reactions rates $\Gamma_\x$ and the resistivity coefficients $\mathcal R^{\x\y}$ are provided by the microphysics, we  have \cite{variate}
\be
R^\x_a =  \Gamma_\x \tilde \mu_\x u^\x_a +  \sum_{\y\neq\x} \mathcal R^{\x\y} (\delta_a^b + v_\x^b u_a)  w^{\y\x}_b  \ ,
\label{pheno}
\ee
for all material particles. The construction is closed  by the constraint
\be
R^\s_a = - \sum_{\x\neq\s} R^\x_a \ ,
\label{Rscon}
\ee
which means that 
\begin{multline}
T \Gamma_\s \approx - u_\s^a R^\s_a = (u^a + v_\s^a) \sum_{\x\neq\s} R^\x_a
\approx \ - \sum_{\x\neq\s} \left[ \Gamma_\x \tilde \mu_\x  + \sum_{\y\neq\x} \mathcal R^{\x\y} w_{\x\s}^a w^{\y\x}_b\right] \\
=   \Gamma_\e \beta +   \sum_{\x\neq\s} \sum_{\y\neq\x} \mathcal R^{\x\y} w_{\s\x}^a w^{\y\x}_b \ge 0 \ .
\label{TGs}
\end{multline}
As discussed in \cite{variate}, it follows that the $\mathcal R^{\x\y}$ coefficients are required to be positive by the second law of thermodynamics (they are also symmetric in $\x$ and $\y$).

\subsection{The individual momentum equations}

The vorticity contribution to each of the momentum equations expands to
\begin{multline}
n_\x \mu_\x \left[ u^a \nabla_a u_b + v_\x^a \nabla_a u_b + u^a \nabla_a v^\x_b
+\perp^a_{\x\ b} {1\over \mu_\x}  \nabla_a \mu_\x \right] + 2 n_\x u^a \nabla_{[a} \sum_{\y \neq\x} n_\y \A^{\x\y} w^{\y\x}_{b]}\\
=   e_\x n_\x \left[ e_b + \epsilon_{bac} v_\x^a b^c + u_b \left( v_\x^a e_a\right)\right]  + \Gamma_\x (\tilde \mu_\x u^\x_b - \tilde \mu^\x_b) + R^\x_b \ ,
\label{eq:vorticitycontr}
\end{multline}
where
\be
\perp_\x^{ab} = g^{ab} + u_\x^a u_\x^b \ ,
\ee
and the electric and magnetic fields $e^a$ and $b^a$ (we use lower case letters for the fields measured in the fluid frame to distinguish from the corresponding fields in the Eulerian frame considered in \cite{foliate}) follow from
\be
F_{ab} = 2 u_{[a} e_{b]} + \epsilon_{abcd} u^c b^d  \ ,
\ee
where we will use the shorthand notation $\epsilon_{abd} =  \epsilon_{cabd} u^c$ from now on~\footnote{It is important to note the sign convention here. With our chosen sign a coordinate frame moving along with $u^a$ is given by the usual ``right-handed'' coordinate system.}.  The evolution of the fields $e^a$ and $b^a$ follow from the standard Maxwell equations (see for example \cite{namhd} or  the Appendix in \cite{foliate}). We will not discuss those equations in detail here as they remain unchanged in the multifluid description. We simply assume that the Maxwell part of the problem, coupled to the charge current $j^a$ from the fluid components, can be solved to provide the full Faraday tensor. 

In order to account for all dynamical degrees of freedom in the problem, it is sufficient to work with the projection of the momentum equations orthogonal to $u^b$. 
For the material particles, we need 
\begin{multline}
n_\x \mu_\x \left[ (u^a +v_\x^a) \nabla_a u_b  +\perp^c_b u^a \nabla_a v^\x_c
+\left( \perp^a_{\ b} +v_b^\x u^a\right) {1\over \mu_\x}  \nabla_a \mu_\x \right] + 2 n_\x u^a \nabla_{[a} \sum_{\y \neq\x} n_\y \A^{\x\y} w^{\y\x}_{b]}\\
=  e_\x n_\x \left[ e_b + \epsilon_{bac} v_\x^a b^c \right] + \sum_{\y\neq\x} \left(\mathcal R^{\x\y}- \Gamma_\x  n_\y \A^{\x\y} \right) w_b^{\y\x}
- e_\x \Gamma_\x ( \perp^a_b+ v^\x_b u^a) A_a \ .
\label{orthind}
\end{multline}
Note that the entropy component differs somewhat since the resistivity is then obtained from \eqref{Rscon}; the appropriate form is provided later.
Note also that, at this point it is common to introduce spatially projected derivatives. We will not do so here, as our main focus is on formal aspects of the problem. We refer the interested reader to the discussion in the companion paper \cite{foliate}.

The problem we are considering --- a hot $\n\p\e$-plasma --- is, in general, associated with four distinct flows. In choosing to work in the centre of momentum frame associated with $u^a$ (which corresponds to the  center of mass in the Newtonian case), we have fixed one of the degrees of freedom. To make progress we need to make choices for the remaining three. Ideally, we would like to make choices that help our intuitive understanding. For example, it may be natural to use a weighted difference of the proton-electron momentum equations as this leads to an equation that generalises Ohm's law for the charge current. As in \cite{namhd} this equation follows if we first divide each momentum equation with $n_\x\mu_\x$ and then take the difference. The motivation for the weighting is obvious from \eqref{eq:vorticitycontr}; we remove the explicit presence of the four acceleration from the combined equation. In the spirit of this argument, it may be tempting to use a similar difference, say between the entropy and the electrons, for the heat flow. However, the equation we arrive at would be counter-intuitive as it would explicitly link the heat flux to the Lorentz force acting on the electrons. To avoid confusion, it may be better to work directly with the entropy momentum equation. Of course, this does not actually remove the coupling to the electromagnetic field. The coupling is just not as explicit. A similar argument applies to the neutrons. As these examples indicate, it is not clear that there is a ``best'' choice of equations for this complex problem. Different choices may be preferred in different situations. With this in mind, we opt to work with the three individual momentum equations for the neutrons $v_\n^a$, the entropy $q^a$ and the electrons, where $v_\e^a$ [once we make use of the frame condition \eqref{eq:observer}] acts as a proxy for the charge current $J^a$.

\subsection{Energy/momentum conservation}

The motion of the zero-momentum flux observer is determined by $\nabla_a T^{ab}=0$ for the total stress-energy tensor. To linear order (in the relative velocities) we get
\be
\nabla^a T_{ab}^\mathrm{M}= g_{ab}\nabla^a p +(p+\varepsilon)u_a \nabla^a u_b + u_b \nabla^a[(p+\varepsilon)u_a]
= - J^a F_{ab} \ . \label{eq:tab} 
\ee
This leads to the usual  equations for energy and momentum conservation;
\be
u^a\nabla_a \varepsilon + (p+\varepsilon) \nabla_a u^a = J^a e_a \ ,
\label{entot}\ee
and 
\be
(p+\varepsilon) u^a\nabla_a u_b + \perp^a_{\ b} \nabla_a p = \epsilon_{bac} J^a b^c \ ,
\label{momtot}\ee

As an alternative to evolving the energy, we may opt to work directly with the entropy. We then need
\be
\nabla_a s^a = \Gamma_\s \ge 0 \ ,
\ee
in accordance with the second law of thermodynamics. When the drift velocity relative to the chosen frame is small, this leads to
\be
u^a \nabla_a s + s \nabla_a u^a + \nabla_a \left( {q^a \over T} \right)  = \Gamma_\s \ ,
\label{GamS2}
\ee
which is completed by the entropy rate from \eqref{TGs}.

\subsection{Ohm's law}

When different components are decoupled from the bulk flow, we need to consider  additional degrees of freedom. Once we have decided which variables to work with, here $J^a$, $q^a$ and $v_\n^a$, we can readily write down the relevant momentum equations that follow from  \eqref{orthind}. Starting with the electron momentum [and making use of \eqref{momtot}] we have
\begin{multline}
e n_\e \mathcal E_b - \left( 1 - {n_\e \mu_\e \over p+\varepsilon}\right) \epsilon_{bac} J^a b^c - {1\over n_\e e} \left( \hat {\mathcal R} - \Gamma_\e s \A^{\e\s} \right) J_b   \\
= - en_\e \epsilon_{bac} v_\p^a b^c + \mathcal R_{\e\n} w^{\n\p}_b + \left( \mathcal R_{\e\s} - \Gamma_\e s \A^{\e\s} \right) \left( {q_b \over sT} - v^\p_b\right)  \\
-n_\e \mu_\e \left[ \left( v_\p^a - {J^a\over e n_\e} \right) \nabla_a u_b + \perp^c_b u^a \nabla_a \left( v^\p_c - {J_c\over e n_\e} \right)
+ \left( v^\p_b - {J_b \over e n_\e} \right) u^a {1\over \mu_\e} \nabla_a \mu_\e \right] \\
+ 2 n_\e u^a \nabla_{[a} s\A^{\e\s} w^{\s\e}_{b]} - e \Gamma_\e \left[ \perp^a_b+ u^a \left( v^\p_b -{J_b\over en_\e}\right) \right] A_a \ ,
\label{ohmfinal}
\end{multline}
where we have introduced the electro-chemical field \cite{bland}
\be
\mathcal E^a = e^a + {1\over e} \perp^{ab} \left(  \nabla_b \mu_\e - {\mu_\e \over p+\varepsilon} \nabla_b p \right) \ ,
\label{electrochem}
\ee
and the total resistivity affecting the electrons;
$\hat{\mathcal R} = \mathcal R_{\e\p} + \mathcal R_{\e\n} +  \mathcal R_{\e\s}$. We have also assumed that  the electrons may entrain the entropy, which would lead to $\mathcal A^{\e\s}\neq 0$. This is the only entrainment coupling that enters the electron momentum equation.
Note also that, we have chosen to write the equation in this particular way because the left-hand side does not change when we consider the problem in different useful limits, e.g. when various components are coupled. 

The equation is, of course, not yet complete. We also have the frame condition \eqref{eq:observer}. In the general case, when we consider all relative flows, this leads to 
\be
v_\p^a =-   X_\n v_\n^a + { X_\e J^a \over n_\e e} - { X_\s \over sT} q^a \ ,
\ee
where we have introduced the dimensionless ``weighting'' factors;
\be
X_\n = {n_\n \mu_\n \over n_\e (\mu_\p+ \mu_\e)} \ , \qquad X_\e = {\mu_\e \over \mu_\p + \mu_\e} \ , \qquad X_\s = {sT \over n_\e ( \mu_\p + \mu_\e)} \ .
\ee
It follows that 
\be
w_{\p\n}^a = - ( 1+ X_\n) v_\n^a + { X_\e J^a \over n_\e e} - { X_\s \over sT} q^a \ ,
\ee
and
\be
w_{\s\e}^a =  X_\n v_\n^a + (1- X_\e) {J^a \over n_\e e} + (1+  X_\s) {q^a \over sT} \ ,
\ee
and it is straightforward to express the equations in our chosen variables. 

The problem is clearly complicated, but we can identify key features. In particular,  it is worth highlighting the role of the four acceleration  in the various momentum equations. In \eqref{ohmfinal} the use of \eqref{momtot} led to the second term in the combination;
\be
\nabla_a \mu_\e - {\mu_\e \over p+\varepsilon} \nabla_a p \ ,
\ee
in the electro-chemical potential.
In a dynamical setting the actual meaning of this combination may not be obvious, but if we consider a static star (say) we have a clear interpretation. For a static star, with metric such that $g_{tt} = -e^\nu$, we have the hydrostatic equilibrium equation
\be
p'=-{1\over 2} (p+\varepsilon) \nu' \ ,
\ee
with the primes denoting radial derivatives. In this case, it follows that
\be
\nabla_a \mu_\e - {\mu_\e \over p+\varepsilon} \nabla_a p = e^{-\nu/2} {d\over dr} \left( \mu_\e e^{\nu/2} \right) =  e^{-\nu/2} {d\over dr} \left( \mu_\e^\infty  \right) \ .
\ee
We learn that the pressure gradient  encodes the gravitational redshift of the ``energy'' term $\mu_\e$ (the same will be true for $T$ and $\mu_\n$ later).

In order to gain further confidence, it is useful to consider limiting cases where different components are strongly coupled, e.g. due to a dominant inter-component friction. Such models follow readily from the general case. We only need to redefine the weighting factors (and cross out  ``undesired'' terms).

As an example, let us  consider the case where only the electrons are free to move relative to the other components, i.e. when we only have the charge current. Then $w_{\p\n}^a = w_{\s\p}^a = 0$ and the frame condition leads to
\be
v_\p^a =    { Y_\e J^a \over n_\e e} \ , \qquad \mbox{with} \qquad Y_\e = {n_\e \mu_\e \over p + \varepsilon} \ .
\ee
Using these results in Ohm's law \eqref{ohmfinal}, we arrive at
\begin{multline}
e n_\e \mathcal E_b - \left( 1 - 2Y_\e \right) \epsilon_{bac} J^a b^c - {1\over n_\e e} \left( \hat {\mathcal R} - \Gamma_\e s \A^{\e\s} \right) J_b \\
= 
n_\e \mu_\e \left[ (1-Y_\e) {J^a\over e n_\e} \nabla_a u_b + \perp^c_b u^a \nabla_a \left[ (1-Y_\e) {J_c\over e n_\e}  \right]
+ \left(1-Y_\e \right) {J_b \over e n_\e \mu_\e } u^a  \nabla_a \mu_\e \right] \\
- 2 n_\e u^a \nabla_{[a} {s\A^{\e\s} \over en_\e} J_{b]} - e \Gamma_\e \left[ \perp^a_b- u^a (1-Y_\e){J_b\over en_\e}\right] A_a \ .
\end{multline}

In the non-relativistic limit, we have $Y_\e \ll 1$  since $m_\e \ll m_\b$, the baryon mass. If we take this limit and  ignore reactions, our final result  reduces to
\be
e n_\e \mathcal E_b - \epsilon_{bac}J^a b^c - \left\{ \hat {\mathcal R} +  {1\over \mu_\e } u^a \nabla_a  \mu_\e \right\}  {J_b \over e n_\e} 
=  n_\e\mu_\e \left[  {J^a\over e n_\e} \nabla_a u_b +\perp^c_b u^a \nabla_a \left( {J_c \over e n_\e} \right)   
\right] \ .
\ee
This agrees with the corresponding limit for the two-component case from [cf. \cite{namhd}, eq (80)]. The neutral bystander enters mainly through the battery term. Note also that the constraint from the second law implies that $\hat {\mathcal R}\ge 0$.

\subsection{The heat equation}

In order to describe the thermal component we need to consider the entropy, which is conveniently described by \eqref{GamS2}.
In addition, the equation for the heat flux follows from the entropy momentum equation. If we account for scattering off of all material particles and also assume that the entropy can be entrained with all particles 
\begin{multline}
\perp^a_b \nabla_a T + T\dot u_b -  {\Gamma_\s \over s^2} q_b + {1\over s} \left(  \perp^a_b \dot q_a + q^a \nabla_a u_b + q_b \nabla_a u^a \right)  \\
= - {1\over s} \sum_{\x\neq\s} \left( \Gamma_\x \tilde \mu_\x v^\x_b + \sum_{\y\neq\x} \mathcal R^{\x\y} w^{\y\x}_b \right) \\
-   \sum_{\x\neq\s} \left( {1\over s} \Gamma_\s n_\x \A^{\x\s} w^{\x\s}_b + 2 u^a\nabla_{[a} n_\x \A^{\x\s}w^{\x\s}_{b]} \right) \ .
\label{heatfinal}
\end{multline}

As in the case of Ohm's law, it is instructive to simplify the problem to two components. Here we do this by assuming that the material particles are locked in \eqref{heatfinal}. Letting the associated drift velocity (relative to the centre of momentum frame) be $v_\b^a$ we then have $v_\b^a = v_\n^a = v_\p^a = v_\e^a$ and 
\be
v_\b^a = - Y_s {q^a \over sT} \ , \qquad \mbox{with} \qquad Y_\s = {sT \over n_\n \mu_\n + n_\e ( \mu_\p + \mu_\e) } \ ,
\ee
and
\be
 w_{\b\s}^a = - (1+ Y_\s) {q^a \over sT} \ .
\ee
Thus we get~\footnote{It is worth noting that, if we use \eqref{momtot} to replace the four acceleration in this expression (in order to introduce the redshifted temperature) then an additional electromagnetic coupling becomes explicit. }
\begin{multline}
\perp^a_b \nabla_a T + T\dot u_b 
= -  {1\over s} \left(  \perp^a_b \dot q_a + q^a \nabla_a u_b + q_b \nabla_a u^a \right) \\
+  \left[  T \Gamma_\s  - \Gamma_\e \beta Y_\s -  (1+ Y_\s) \sum_{\x\neq\s}  \left( \mathcal R^{\s\x}  + \Gamma_\s  n_\x \A^{\x\s} \right)   \right] {q_b \over s^2 T} +  \\
+ 2  \sum_{\x\neq\s} u^a\nabla_{[a} (1+ Y_\s)   {n_\x \A^{\x\s} \over sT} q_{b]}  \ ,
\end{multline}
or, if we assume that $Y_\s\ll 1$;
\begin{multline}
\perp^a_b \nabla_a T + T\dot u_b 
= -  {1\over s} \left(  \perp^a_b \dot q_a + q^a \nabla_a u_b + q_b \nabla_a u^a \right) \\
- {1\over s} \left[  \sum_{\x\neq\s}  \left( \mathcal R^{\s\x}  + \Gamma_\s  n_\x \A^{\x\s} \right) +  \Gamma_\e \beta Y_\s - T \Gamma_\s     \right] {q_b \over sT}  
+ 2  \sum_{\x\neq\s} u^a\nabla_{[a}    {n_\x \A^{\x\s} \over sT} q_{b]}   \ .
\label{heats}
\end{multline}

It is useful to explain how this reduces to the more familiar form for the heat equation. Let us ignore reactions and the entrainment, even though we know that it may be important to ensure causality and stability \cite{heat1,heat2}. Then, noting that the resistivity leads to $\Gamma_\s$ being quadratic in the drift velocities, we see that it is natural to define the heat conductivity as
\be
\kappa =  {s^2 T} \left[  \sum_{\x\neq\s} \mathcal R^{\s\x}   \right]^{-1} \ ,
\ee
 and a relaxation time $\tau = {\kappa/ s}$ to get
\be
 \tau  \left( \perp^a_b \dot q_a + q^a \nabla_a u_b + q_b \nabla_a u^a \right) + q_b = - \kappa \left( \perp^a_b \nabla_a T + T\dot u_b  \right) \ .
\ee

This equation is coupled to the entropy equation \eqref{GamS2}, which for this particular model problem takes the form (as we ignore particle reactions)
\be
u^a \nabla_a s + s \nabla_a u^a + \nabla_a \left( {q^a \over T} \right) = {q^2 \over \kappa T^2} \ .
\ee

\subsection{Decoupling superfluid neutrons}

As a final two-fluid example,  let us consider the decoupling of superfluid neutrons. In principle, the onset of superfluidity will suppress both nuclear reactions and resistive scattering, but let us nevertheless keep the relevant terms (in the first instance). In addition, let us assume that the neutrons entrain protons (due to the strong interaction) and the entropy.
With these assumptions, the neutron momentum equation takes the form
\begin{multline}
n_\n \mu_\n \left[  \perp^c_b\dot v^\n_c +  v_\n^a \nabla_a u_b \right]  + n_\n v^\n_b \dot \mu_n + n_\n \perp^a_b \left( \nabla_a \mu_\n - {\mu_\n \over p+\varepsilon} \nabla_a p \right) + {n_\n \mu_\n \over p+\varepsilon} \epsilon_{bac} J^a b^c \\
=  \sum_{\x\neq \n} \mathcal R^{\n\x} w_b^{\x\n} - \sum_{\x=\{\p,\s\}}  \left(  \Gamma_\n  n_\x \A^{\n\x} w_b^{\x\n} + 2n_\n u^a \nabla_{[a} n_\x \A^{\n\x} w^{\x\n}_{b]} \right) \ .
\label{sffinal}
\end{multline}

As in the previous cases, we can help our intuition by locking all components apart from the neutrons. Now we have $v_\b^a=v_\p^a=v_\e^a=v_\s^a$ and
\be
v_\b^a = - Y_\n v_\n^a \ , \qquad \mbox{with} \qquad Y_\n = {n_\n \mu_\n \over n_\e (\mu_\p+\mu_\e) + sT}  \ ,
\ee
so
\be
w_{\b\n}^a = - (1+Y_\n) v_\n^a \ ,
\ee
and since we must have $J^a=0$ in this case, we see that \eqref{sffinal} leads to
\begin{multline}
n_\n \mu_\n \left[  \perp^c_b\dot v^\n_c +  v_\n^a \nabla_a u_b \right]  + n_\n v^\n_b \dot \mu_\n + n_\n \perp^a_b \left( \nabla_a \mu_\n - {\mu_\n \over p+\varepsilon} \nabla_a p \right) \\
=- \left[ \sum_{\x\neq\n} \mathcal R^{\n\x}  - \Gamma_\n ( n_\p \A^{\n\p} +s \A^{\n\s}) \right](1+ Y_\n) v^\n_b \\
+ 2n_\n u^a \nabla_{[a} (1+ Y_\n) ( n_\p \A^{\n\p} +s \A^{\n\s})v^{\n}_{b]} \ .
\end{multline}

For a strongly superfluid system, sufficiently cold that we can ignore the thermal component, there will be no scattering involving neutrons and  reactions are suppressed. Then we have 
(noting that it would not be reasonable to assume that $Y_\n \ll 1$);
\begin{multline}
n_\n \mu_\n \left[  \perp^c_b\dot v^\n_c +  v_\n^a \nabla_a u_b \right]  + n_\n v^\n_b \dot \mu_\n + n_\n \perp^a_b \left( \nabla_a \mu_\n - {\mu_\n \over p+\varepsilon} \nabla_a p \right) \\
- 2n_\n u^a \nabla_{[a} (1+ Y_\n)  n_\p \A^{\n\p} v^{\n}_{b]} = 0 \ ,
\end{multline}
or
\be
  \perp^c_b \left\{ u^a \nabla_a (m_\n^\star v^\n_c )  +  \left( \nabla_c \mu_\n - {\mu_\n \over p+\varepsilon} \nabla_c p \right)
+  2 \mu_\n v_\n^a \nabla_{[a} u_{c]} +m_\n^\star v_\n^a \nabla_c u_a \right\} = 0 \ ,
\ee
where we have introduced the effective neutron mass in the centre of momentum frame;
\be
m_\n^\star = \mu_\n - (1+Y_\n) n_\p \A^{\n\p} \ .
\ee
This result basically shows that, when there is no rotation or shear associated with $u^a$, and if the composition is uniform, then the superfluid 
flow is potential. It is straightforward to sanity check this result because the standard two-fluid model, which involves working in the frame of the ``normal'' component (see, for example, \cite{livrev}), is obtained by setting $Y_\n=0$. This demonstrates that it is straightforward to change the observer frame in the formalism.

\subsection{A three-fluid model: The thermo-electric effect}

Stepping up the level of complexity, we can write down different three-fluid models by assuming that only two of the four components are locked. Let us focus on one of the possibilities, with both charge current and heat flow. This example is interesting because it introduces the thermo-electric effect, which may be relevant for the neutron star magnetic field evolution \cite{bland}. 

To arrive at this model we keep neutrons and protons locked, but allow both electrons and heat to flow. The frame choice then leads to
\be
v_\p^a = \tilde Y_\e {J^a \over en_\e} - \tilde Y_\s {q^a \over sT} \ ,
\ee
where 
\be
\tilde Y_\e = {n_\e \mu_\e \over n_\n \mu_\n + n_\e (\mu_\p + \mu_\e)} \ , \qquad  \tilde Y_\s = {sT \over n_\n \mu_\n + n_\e (\mu_\p + \mu_\e)} \ .
\ee
The electron momentum equation now leads to a more complicated Ohm's law. If we ignore nuclear reactions, then this equation takes the form 
\begin{multline}
e n_\e \mathcal E_b - \left(1-\tilde  Y_\e - {n_\e \mu_\e \over p+\varepsilon} \right) \epsilon_{bac}J^a b^c - \left\{ \hat  {\mathcal R} +  {1\over \mu_\e } u^a \nabla_a [( 1- \tilde  Y_\e) \mu_\e] \right\}  {J_b \over e n_\e} \\
=  n_\e\mu_\e \left[  (1-\tilde  Y_\e) {J^a\over e n_\e} \nabla_a u_b + (1-\tilde  Y_\e) \perp^c_b u^a \nabla_a \left( {J_c \over e n_\e} \right)    
\right] \\
+ {e n_\e \over sT} \tilde  Y_\s \epsilon_{bac} q^a b^c + (1+\tilde  Y_\s)  \mathcal R^{\e\s} {q_b \over sT}
\\
+ {n_\e \mu_\e \over sT} \tilde  Y_\s  \left( q^a \nabla_a u_b  + \perp^c_b u^a \nabla_a q_c \right) 
+ n_\e q_b u^a \nabla_a \left( {\mu_\e \tilde  Y_\s \over sT} \right) \\
- n_\e u^a \nabla_a \left\{ s\A^{\e\s} \left[ (1+\tilde  Y_\s) { q_b \over sT} + (1-\tilde  Y_\e) {J_b \over e n_\e} \right]\right\} \\
- n_\e s\A^{\e\s} \left[ (1+\tilde  Y_\s) { q^a \over sT} + (1-\tilde  Y_\e) {J^a \over e n_\e} \right] \nabla_b u_a \ ,
\end{multline}
where we have made use of \eqref{momtot} and (re)-defined $\hat {\mathcal R} = \mathcal R^{\e\n} + \mathcal R^{\e\p} +(1-\tilde Y_\e) \mathcal R^{\e\s}$.

This result is not very transparent so let us  assume the  $\tilde Y_\e \ll 1$ and $\tilde Y_\s \ll 1$, as before. If we also ignore the entropy entrainment, we  have 
\begin{multline}
e n_\e \mathcal E_b - \epsilon_{bac}J^a b^c - \left( \hat {\mathcal R} +  {1\over \mu_\e } u^a \nabla_a  \mu_\e \right) {J_b \over e n_\e} 
 \\
= n_\e\mu_\e \left[   {J^a\over e n_\e} \nabla_a u_b + \perp^c_b u^a \nabla_a \left( {J_c \over e n_\e} \right)    
\right] \\
+ {e n_\e \over sT} \tilde Y_\s \epsilon_{bac} q^a b^c +  \mathcal R^{\e\s}  {q_b \over sT}
\\
+ {n_\e \mu_\e \over sT} \tilde Y_\s \left( q^a \nabla_a u_b  + \perp^c_b u^a \nabla_a q_c \right) 
+ n_\e q_b u^a \nabla_a \left( {\mu_\e \tilde Y_\s \over sT} \right)  \ .
\label{thermo1}
\end{multline}

Meanwhile, the corresponding heat equation becomes (noting that, in absence of reactions, the term involving $\Gamma_\s$ is quadratic in the drift velocities \cite{variate}) 
\be
\perp^a_b \nabla_a T + T\dot u_b  + {1\over s} \left(  \perp^a_b \dot q_a + q^a \nabla_a u_b + q_b \nabla_a u^a \right)  \\
= -    \mathcal R^{\s\e}  {J_b \over e s n_\e} - {1\over s^2 T}\left(  \sum_{\x\neq \s} \mathcal R^{\x\s} \right)  q_b \ .
\label{thermo2}
\ee

As we are mainly interested in seeing how the heat flow couples to the charge current, we note that \eqref{thermo1} may be approximated as 
\be
e n_\e \mathcal E_b - \epsilon_{bac}J^a b^c - \hat {\mathcal R} {J_b \over e n_\e} \approx   \mathcal R^{\e\s}  {q_b \over sT}  = {s\over \kappa} q_b \ ,
\ee
where we have assumed  that $\mathcal R^{\e\s}$ makes the dominant contribution to the thermal conductivity, so we have
\be
\kappa \approx  {s^2 T \over \mathcal R^{\e\s}} \ .
\ee
Meanwhile \eqref{thermo2} leads to 
\begin{multline}
q_b \approx - \kappa( \perp^a_b \nabla_a T + T\dot u_b )- { sT \over e n_\e} J_b \\
= - \kappa \perp^a_b \left( \nabla_a T - {T\over p+\varepsilon} \nabla_a p\right) - {sT \over e n_\e} J_b - {\kappa T \over p+\varepsilon} \epsilon_{bac} J^a b^c \ .
\end{multline}
Combining the two relations we have
\be
e n_\e \mathcal E_b + s \perp^a_b \left(\nabla_a T - {T\over p+\varepsilon} \nabla_a p \right) = S_{ba}J^a \ ,
\ee
where
\be
S_{ab} \approx {1\over e n_\e}  \left( \mathcal R^{\e\n} + \mathcal R^{\e\p} \right) \perp_{ab} + \epsilon_{abc} b^c \ .
\ee
Writing this as 
\be
S_{ab} = {1\over \sigma}  \left(  \perp_{ab} +\zeta  \epsilon_{abc} b^c \right) \ ,
\ee
we have the inverse \cite{namhd}
\be
\sigma^{ab} = {\sigma \over \zeta^2 b^2} \left( \perp^{ab} + \zeta^2 b^a b^b - \zeta \epsilon^{abc} b_c\right)   \ ,
\label{inve}\ee
and the final relation
\be
J^a = e n_\e \sigma^{ab}\mathcal E_b + s \sigma^{ab} \perp^c_b  \left( \nabla_c T - {T\over p+\varepsilon} \nabla_c p \right) \ .\label{thermo3}
\ee
The final term  encodes the thermo-electric effect. 

\section{Concluding remarks}

In this paper we have outlined a new state-of-the-art for the modelling of charged fluid dynamics in general relativity. Our framework can be applied to a wide range of problems in relativistic astrophysics and cosmology, ranging from the secular magneto-thermal evolution of neutron stars to the violent dynamics of neutron star mergers and various explosive scenarios to the large scale evolution of the magnetic universe. 

After outlining the general framework (building on \cite{livrev, monster} and recent progress from \cite{variate}) we discussed the connection with standard relativistic magnetohydrodynamics. This discussion highlighted the need to go beyond the usual description in order to account for relevant effects like the inertia of both charge currents and heat and, for mature systems, the decoupling of superfluid components. Restricting ourselves to systems such that the relative flow relative to a chosen ``fluid'' observer are small, we discussed the connection with the microphysics and the equation of state. We also introduced a phenomenological resistivity in order to account for a range of friction mechanisms. 

In order to illustrate the use of the new formalism in various settings we considered three distinct two-fluid models describing i) an Ohm's law for resistive charged flows, ii) a relativistic heat equation, and iii) an equation representing the momentum of a decoupled superfluid component. In each case, we considered how the results connect with previous results in the literature. These demonstrations provide confidence in the model. Of course, the actual aim is higher. As a more complex example, we used the framework to formulate a three-fluid model demonstrating the thermo-electric effect. In principle, one would expect this effect to operate in  young neutron stars, but as soon as the star becomes isothermal --- after 1,000 years or so --- the coupling between the charge current and the heat flow will be quenched. However, at this point the star's core will have become  superfluid. This brings another coupling mechanism into play. 
Given our framework, it is easy to write down a model that decouples the superfluid neutrons (and either ignores the thermal component or locks it to the charged components). In this model, the flow of the superfluid couples to the charge current and  impacts on the magnetic field evolution. Analogously, we obtain a third three-fluid model by ignoring the charge current (locking protons and electrons). This model is interesting because it shows how the superfluid affects the heat flow. This model connects with the discussion of the role of the superfluid phonons for heat conduction \cite{reddy}.

 In essence, the flexible framework we have developed allows us to model neutron stars (and related systems) at a hierarchy of increasingly complex levels. This should enable us to make interesting progress on a range of  problems. As a first step towards related numerical simulations we develop the 3+1 spacetime foliation view of the problem in a companion paper \cite{foliate}. A set of related two-fluid plasma simulations are described in \cite{kiki}.
 
 \acknowledgments

NA, IH and KD gratefully acknowledge support from the STFC.

\end{document}